\newif\ifuseprd
\newif\ifeprint
\newif\ifdatelast

\useprdtrue
\eprinttrue
\datelastfalse

\documentclass[preprint,
tightenlines,%
floats,prd,eqsecnum,nobibnotes%
,nofootinbib,aps,12pt]{revtex4}
\usepackage{amsmath,amssymb}
\usepackage{graphicx}
\let\oldappendix\appendix
\renewcommand\appendix{\oldappendix%
    \renewcommand\theequation{\thesection.\arabic{equation}}}
\providecommand\FIGURE[2][]{\begin{figure}[#1]\begin{center}{#2}\end{center}
                       \end{figure}}

\allowdisplaybreaks[3]

\newif\iftoomuchdetail
\toomuchdetailtrue
\toomuchdetailfalse
\newcounter{saveequation}
\newcounter{detailnum}
\newcommand\savetheequation{\theequation}
\newcommand\detailtheequation{%
	  $\delta$\Roman{detailnum}:\roman{equation}}
\newenvironment{detail}{\iftoomuchdetail\sf
         \setcounter{saveequation}{\value{equation}}%
         \setcounter{equation}{0}\addtocounter{detailnum}{1}%
         \renewcommand\theequation\detailtheequation%
         \fi}{
     \iftoomuchdetail%
     \ifnum\value{equation}=0\addtocounter{detailnum}{-1}\fi%
     \setcounter{equation}{\value{saveequation}}%
     \renewcommand\theequation\savetheequation%
     \fi%
     }

\newcommand\abs[1]{\ensuremath{\left\lvert{#1}\right\rvert}}

\newcommand\com[2]{\ensuremath{\left[{#1},{#2}\right]}}
\newcommand\ket[1]{\ensuremath{\lvert{#1}\rangle}}
\newcommand\bra[1]{\ensuremath{\langle{#1}\rvert}}
\DeclareMathOperator{\Tr}{Tr}
\newcommand\mathone{{\rlap{\kern .25em l}1}}
\newcommand\one{{\ifmmode{\text{\mathone}}\else{\mathone}\fi}}

\newcommand\field[1]{{\ensuremath{\mathbb{{#1}}}}}
\newcommand\ZZ{{\field{Z}}}

\newcommand\ZR{{\field{R}}}

\newcommand\ppwave{{\em pp}~wave}

\makeatletter
\def\@strike{\relax\leavevmode
  \ifmmode
    \expandafter\mathpalette\expandafter\math@strike
  \else
    \expandafter\make@strike
  \fi}
\def\math@strike#1#2{%
  \setbox\z@\hbox{$\m@th#1{#2}$}\fin@strike}
\def\make@strike#1{%
  \setbox\z@\hbox{\color@begingroup#1\color@endgroup}\fin@strike}
\def\fin@strike{%
  \@tempdima\dp\z@
  \@tempdimb\ht\z@
  \lower\@tempdima\hbox{\strike@start}%
  \box\z@
  \raise\@tempdimb\hbox{\strike@end}}
\def\strike@start{\special{ps: %
    currentpoint /starty exch def /startx exch def}}
\def\strike@end{
}
\newcommand\fs{\protect\@strike}

\newcommand\citejournal[4]{{\ifuseprd\else\begingroup\em\fi {#4}%
     \ifuseprd\else\endgroup\fi {\bf {#1}}\ifdatelast, {#3} ({#2})\else%
     \ ({#2}) {#3}\fi}}

\providecommand\plb[3]{{\citejournal{#1}{#2}{#3}{Phys.\ Lett.\ B }}}
\providecommand\npb[3]{{\citejournal{{\ifuseprd\bf\else\em\fi B\/}#1}{#2}{#3}{Nucl.\ Phys.\ }}}
\providecommand\jhep[3]{{\citejournal{#1}{#2}{#3}{J.\ High Energy Phys.\ }}}
\providecommand\npps[3]{{\citejournal{}{#2}{#3}{Nucl.\ Phys.\  Proc.\ Suppl.\ %
     {\normalfont\bf {#1}}}}}

\providecommand\cqg[3]{{\citejournal{#1}{#2}{#3}{Class.\ Quant.\ Grav.\ }}}
\providecommand\mpla[3]{{\citejournal{{\em A\/}#1}{#2}{#3}{Mod.\ Phys.\ %
   Lett.\ }}}

\providecommand\citeprd[3]{{\citejournal{#1}{#2}{#3}{Phys.\ Rev.\ D }}}

\providecommand\citeprl[3]{{\citejournal{#1}{#2}{#3}{Phys.\ Rev.\ Lett.\ }}}

\providecommand\hepth[1]{{\ifuseprd{\eprint{{\ifeprint\tt\fi hep-th/#1}}}%
                \else{\tt hep-th/{#1}}\fi}}

\newcommand\phepth[1]{{\ifuseprd\else\tt\fi [\hepth{#1}]}}
\newcommand\ct[1]{{\ifeprint\ifuseprd{\em{#1}},\else{\sf {#1}},\fi\fi}}
\newcommand\bt[1]{{\em {#1}},}

\newcommand\ben{\begin{equation}}
\newcommand\een{\end{equation}}
\newcommand\bea{\begin{eqnarray}}
\newcommand\eea{\end{eqnarray}}

\newcommand\nn{\nonumber}

\newcommand\ie{{\em i.e.\/}}
\newcommand\cf{{\em cf.\/}}

\newcommand\etal{{\em et.\ al.\/}}

\begin{document}

\ifeprint
\setlength{\baselineskip}{1.2\baselineskip} 
\fi

\title{\vspace*{\fill}{\LARGE Matrix Membrane Big Bangs and D-brane Production}}
\author{Sumit R. Das}\email{das@pa,uky,edu}
\author{Jeremy Michelson}\email{jeremy@pa,uky,edu}
\affiliation{Department of Physics and Astronomy \\
      University of Kentucky,\\
      Lexington, KY \ 40506 \ U.S.A.
      \vspace*{\fill}}

\begin{abstract}
\vspace*{\baselineskip}
We construct Matrix Membrane theory in \ppwave\ 
backgrounds that have a null linear dilaton
in Type IIB string theory. Such backgrounds can serve as toy models of
big bang cosmologies. 
At late times only abelian degrees of freedom survive, 
and if the Kaluza-Klein modes along one of the directions of the membrane
decouple, standard perturbative strings emerge. Near the ``big
bang'', non-abelian configurations of fuzzy {\em ellipsoids} are present, as in
the Type IIA theories. A generic configuration of these shrink to zero
volume at late times. However, the Kaluza Klein modes
(which can be thought of as states of $(p,q)$ strings 
in the original IIB theory) can be generically produced in pairs
in both \ppwave\
and flat backgrounds in the presence of time dependence.
Indeed, if we require that at
late times the theory evolves to the perturbative string vacuum,  
these modes must be
prepared in a squeezed state with a thermal distribution at 
early times.
\vspace*{\fill}
\end{abstract}

\preprint{\parbox[t]{10em}{\begin{flushright}
UK/06-02 \\ {\tt hep-th/0602099}\end{flushright}}}

\maketitle

\ifeprint
\tableofcontents
\fi

\section{Introduction}\label{intro}

It has been always difficult to address questions involving strongly
time dependent backgrounds in string theory, particularly those which
involve an apparent ``beginning'' or ``end'' of time.  Recently there have
been several attempts to investigate such backgrounds which have
holographic duals in the form of open string theories. These include
nontrivial solutions corresponding to closed string tachyon
condensation of two dimensional non-critical string theory
\cite{Das:2004aq, Das:2005jp}; certain simple backgrounds of
critical string theory which admit a Matrix Theory or a Matrix String
Theory formulation~\cite{Craps:2005wd,Li:2005sz,Li:2005ti,Hikida:2005ec,%
Das:2005vd,Chen:2005mg,She:2005mt,Chen:2005bk,Ishino:2005ru,Robbins:2005ua,%
KalyanaRama:2005uw,Hikida:2005xa}; as well
as models with tachyon condensation in critical string theory~\cite{%
McGreevy:2005ci,Silverstein:2005qf,Berkooz:2005ym,Pawlowski:2005bs,%
Kawai:2005jx,Moore:2005wp,McInnes:2005su,Ross:2005ms,Bergman:2005qf,%
Narayan:2005dz}.
The first two sets of examples have a common feature: the
underlying quantum mechanics is always defined in the open string
formulation which  has a ``time'' and this open string time
runs over the full range of values, while the ``time'' which 
{\em emerges} in the closed string interpretation appears to have
a ``beginning'' or an ``end''.

In the Matrix Theory models, the
Yang-Mills theory of D-branes provide a non-perturbative formulation
of string theory in a time dependent dilaton background with some
fixed quantized momentum $N$ along a null direction.  Consider for
example the case where the string theory is
weakly coupled in the future and strongly coupled in the past. Then,
in the holographic Matrix string theory description
(obtained using~\cite{Banks:1996vh,Motl:1997th,Banks:1996my,Dijkgraaf:1997vv}%
\footnote{
It was shown in~\cite{mbh}
that one can obtain a supersymmetric quantum mechanics
via dimensional reduction of ${\mathcal N}=1, D=10$ SYM.
We thank M.~Halpern for bringing this to our attention.}
and the arguments of
\cite{Sen:1997we,Seiberg:1997ad}) the reverse happens---i.e.\
the $SU(N)$ Yang-Mills coupling is strong in the future and weak
in the past. This means that at late times, the fields of the
Yang-Mills theory are constrained to lie in the Cartan subalgebra and
become the coordinates of the emerging ``space''---the Yang-Mills action
reduces to the worldsheet action of multiple strings (in the light
cone gauge) in this ``space''.  At early times, 
all the non-abelian degrees of freedom are important and there is no
interpretation in terms of strings any more. Equivalently, the
closed string which emerges from the model may be thought to live in
the future quadrant of the Milne universe\cite{Craps:2005wd}%
\footnote{Perturbative strings in Milne universes have been investigated in
\cite{hs,bhkn,cc,s,lms,lms2,tt,ckr,fm,d,cm,r,%
grs,pb,dp,bdpr}}. Thus,
while the time of the open string theory, i.e.\ Matrix theory, runs
over a full range, the time as perceived in the closed string theory
may appear to begin at a finite point.

In \cite{Das:2005vd} we found solutions of Type IIA string theory in a
\ppwave\ background which have a null linear dilaton, and we constructed
Matrix String Theory in this background starting with the BMN Matrix
theory of 
\cite{Berenstein:2002jq},
compatifying~\cite{Michelson:2002wa}, and following~\cite{Dasgupta:2002hx,%
Dasgupta:2002ru,Kim:2002if,Kim:2002zg,Sugiyama:2002tf,%
Hyun:2002wu,Hyun:2002wp,ni,clp2,Das:2003yq} . The presence of a second
length scale (the background flux of the \ppwave) allowed us to
analyze the model in a regime of parameters where a specific class of
non-abelian configurations---fuzzy spheres \cite{Myers:1999ps}---are relevant. 
We looked at
the dynamics of these fuzzy spheres and found that with {\em generic initial
conditions}, these
oscillate in size, but the maximum size vanishes exponentially fast as
the background evolves from the big bang. This leaves only perturbative
closed strings at late times. For large $N$ these fuzzy spheres become 
spherical D2
branes. This model therefore provides a concrete example in which D-branes
proliferate a typical state of the theory at early times and
tame what would appear as an initial singularity from the point of
view of perturbative closed strings. 

In this note, we consider a \ppwave\ background in
IIB string theory with a null linear dilaton, with two compact
directions, one of which is null. In the absence of
a dilaton background this theory has a holographic description
in terms of the large R-charge sector of a 3+1 dimensional
N=4 Yang-Mills theory \cite{Berenstein:2002jq}, or more 
precisely as a version of a quiver gauge theory constructed
along the lines of \cite{Mukhi:2002ck}.
However, now we have a second dual description as well. 
Following the standard
procedure in flat space
\cite{Motl:1997th,Banks:1996my,ss}
a sector of the theory with some given momentum along the
null direction should be dual to a {\em Matrix Membrane Theory}---a
2+1 dimensional Yang-Mills theory on a torus~\cite{Gopakumar:2002dq}. 
This theory was explicitly constructed
in
\cite{Michelson:2004fh,Michelson:2005iib}.
The 2+1 dimensional Yang-Mills theory lives on
a torus, the ratio of the two sides being the IIB string coupling.
In the absence of a background 
dilaton, and at weak string coupling only abelian 
configurations survive. Furthermore, one of the sides
becomes very small and the Kaluza-Klein modes along this direction decouple.
The resulting theory then becomes the lightcone action for a fundamental
IIB string in the \ppwave\ background. 

Here we construct the Matrix Membrane theory in the presence of a linear null
dilaton. At late times this reduces to the worldsheet action of the 
appropriate IIB string. However at early
times, nonabelian configurations are important, in particular 
{\em fuzzy ellipsoids}. In a way similar to \cite{Das:2005vd} these
ellipsoids generically shrink to zero at late times. However now we have a
new phenomenon. The Kaluza-Klein modes, which are states of $(p,q)$ strings in 
the original IIB theory, now become important at early times.
We find that if we require that the state at late times is the vacuum
of perturbative string theory, the initial state must be a squeezed state
of these $(p,q)$ strings with no net winding number. This model therefore
throws light on the question of initial conditions.\footnote{The fact that
models of this type can be used to address issues of initial conditions was
suggested to us by S. Trivedi.}

\section{IIB \ppwave{s} with null dilaton}

The string frame metric,
RR field strengths and the dilaton $\Phi$ are given by
\begin{equation}
\begin{split}
ds^2 & = 2dx^+dx^- - 4\mu^2[(x^1)^2+\cdots(x^6)^2](dx^+)^2 
- 8\mu x^7 dx^8 dx^+ + [(dx^1)^2 + \cdots (dx^8)^2], \\
F_{+1234} & = F_{+5678} = \mu~e^{Qx^+}, \qquad \Phi  =  - Qx^+.
\end{split}
\label{three} \raisetag{\baselineskip}
\end{equation}
It is easy to check that this solves the low energy equations of
motion.  These solve the equations of motion for $\mu = \mu(x^+)$, and
not just constant $\mu$; however, for most of the paper we will
consider just $\mu=\text{constant}$, although we suspect that
$\mu=\mu_0 e^{Q x^+}$ would be a very interesting case, as
in~\cite{Das:2005vd}.  The coordinates $x^-$ and $x^8$ are compact
\ben
x^- \sim x^- + 2\pi R, \qquad x^8 \sim x^8 + 2 \pi R_B.
\label{four}
\een
We will denote the string coupling of this Type IIB theory by $g_B$
and the string length by $l_B$.

T-dualizing along $x^8$ yields a IIA background with a 
string coupling $g_A$ and an $x^8$ radius given by $R_A$,
\ben
g_A = g_B \frac{l_B}{R_B}, \qquad R_A=\frac{l_B^2}{R_B}.
\label{iiaconstants}
\een
This may be further lifted
to M-theory by introducing another compact direction $x^9$. This
M-theory background is given by
\begin{equation}
\begin{split}
ds^2 & = e^{2Qx^+\over 3} \{ 2dx^+dx^- -
4\mu^2[(x^1)^2+\cdots(x^6)^2+ 4 (x^7)^2](dx^+)^2 
+[(dx^1)^2 + \cdots (dx^8)^2] \} \\
& \qquad + e^{-{4Qx^+\over 3}} (dx^9)^2, \\
F_{+789} & = - 4\mu, \qquad F_{+567}  =  8\mu e^{Qx^+}.
\end{split}
\label{five} \raisetag{\baselineskip}
\end{equation}
Again, we could have taken $\mu=\mu(x^+)$ in this background, as for
the dual background~\eqref{three}.
The various factors of $e^{Qx^+}$ may be understood as follows. 
Typically a NS-NS gauge field will
not acquire any such factor, whereas an RR gauge field will
\cite{Das:2005vd}. This is why $F_{+789}$ does not have such
a factor but $F_{+567}$ does.
In terms of the IIB quantities, the radius of the $x^9$ direction
is $R_9$
and the Planck length of the M-theory is $l_p$, where
\ben
R_9 = g_B \frac{l_B^2}{R_B}, \qquad l_p^3 = g_B \frac{l_B^4}{R_B}.
\label{mtheoryconstants}
\een

\section{Matrix Membrane Theory}
Consider a sector of the IIB theory with momentum $p_-=J/R$ along the
$x^-$ direction. If we treat $x^-$ as the Kaluza-Klein direction of
the M-theory background (\ref{five}) we have a IIA theory living on
{\em two} compact directions $x^8$ and $x^9$, with radii $R_A,R_9$
respectively, and a net D0 brane charge equal to $J$. T-dualizing along
$x^8$ and $x^9$ then leads to a IIA theory with D2 brane charge $J$
living on a torus with sides
\ben
{\tilde R}_8 = g_B{l_B^2 \over R}, \qquad
{\tilde R}_9 = {l_B^2 \over R}.
\label{one}
\een
Matrix membrane theory is the $SU(J)$ supersymmetric 
$2+1$ dimensional Yang-Mills theory
of these $J$ D2 branes.
The dimensionful coupling constant of the YM theory is
\ben
G_\text{YM}^2 = \frac{R}{R_9R_A} = \frac{RR_B^2}{g_Bl_B^4}.
\label{two}
\een
In flat space and a constant dilaton, this Matrix Membrane theory was 
constructed in \cite{Motl:1997th,Banks:1996my,ss}.
The $2+1$ Yang-Mills lives on a torus: the ratio of the two radii is
equal to the string coupling of the original theory \cite{sM}.
Therefore, for small
string coupling, the Kaluza-Klein modes for the smaller
circle decouple, and at the same time the potential restricts the
fields to lie in a Cartan subalgebra; the resulting two dimensional
theory becomes the usual light cone superstring after a suitable
dualization\cite{ss} of the gauge field strength.
For time-independent \ppwave{s}, several physical aspects
expected from such a theory were considered in~\cite{Gopakumar:2002dq},
and the theory was explicitly constructed
in~\cite{Michelson:2004fh,Michelson:2005iib}.

The construction of Matrix membrane theory for the background
given in (\ref{five}) follows the procedure of~\cite{Michelson:2004fh,%
Michelson:2005iib}.%
\footnote{Bonelli~\cite{gb} has found this type of
supersymmetric action by dimensional reduction from a deformed ten dimensional
gauge theory.  However,
Kim \etal~\cite{correct}
have shown that, for example, the M-theory \ppwave\ quantum
mechanics follows from dimensional reduction of ${\mathcal N}=4, d=4$ SYM on
an $S^3$.}
The bosonic terms of the action for $J$ 
D0 branes may be
written down following \cite{Myers:1999ps,TV,preTV,ni,clp2}.

The light cone lagrangian of these $J$ D0 branes is given by
\begin{multline}
L = \Tr  \{   \frac{1}{2Rl_p} G^{+-}G_{IJ}D_\tau X^I D_\tau X^J
-\frac{G^{--}l_p}{2G^{+-}R} 
  +  \frac{R}{4G^{+-}l_p^5}G_{IK}G_{JL}[X^I,X^J][X^K,X^L]
\\ - \frac{i}{2l_p^2} A_{+IJ}X^I X^J \},
\label{generalaction}
\end{multline}
where the indices $I,J,K = 1 \cdots 9$ and $\tau$ is a dimensionless 
light cone time. Using a gauge in which the
M theory gauge potentials are given by
\ben
A_{+89} = 4\mu x^7, \qquad A_{+56} = - 8\mu e^{Qx^+} x^7,
\een
the explicit form of the bosonic part of the Matrix theory lagrangian is
\bea
L = \Tr & \{ & \frac{1}{2Rl_p}(D_\tau X^i)^2 + \frac{1}{2Rl_p}
e^{-2Q\tau}(D_\tau X^9)^2 
 -  \frac{2l_p\mu^2}{R}[(X^1)^2 + \cdots (X^6)^2 + 4 (X^7)^2] \nn \\
& + & \frac{R}{2l_p^5} [ X^9, X^i ]^2 +\frac{R}{4l_p^5}
e^{2Q\tau} [ X^j, X^i ]^2 \nn \\ 
& + & \frac{4\mu i}{l_p^2}X^7~[ X^8, X^9 ]
- \frac{8\mu i}{l_p^2}e^{Q\tau} 
X^7~[ X^5, X^6 ] \},
\label{mtheoryaction}
\eea
where the indices $i,j = 1\cdots 8$.

To obtain the Matrix membrane theory we need to substitute
\ben
X^8  \rightarrow  -i R_A D_\rho, \qquad X^9 \rightarrow 
-i R_9 D_\sigma, \qquad
\Tr  \rightarrow  \Tr \int_0^{2\pi}d\sigma \int_0^{2\pi}d\rho,
\een
where $D_\alpha, \alpha = \tau,\sigma, \rho$ 
are covariant derivatives with a $SU(J)$ gauge
connection $A_\alpha$.
We will also rescale the coordinates $\tau,\sigma, \rho$, the
fields $X^a, a=1 \cdots 7$ and $A_\alpha$ as follows
\begin{equation} \label{rescale}
\begin{aligned}
\tau & \rightarrow l_p \tau, &
\sigma & \rightarrow \frac{l_p^3}{RR_9} \sigma, &
\rho & \rightarrow \frac{l_p^3}{RR_A} \rho, \\
X^a & \rightarrow \frac{{\sqrt{RR_AR_9}}}{l_p^3}X^a, &
A_\tau & \rightarrow \frac{1}{l_p} A_\tau, &
A_\sigma &\rightarrow \frac{RR_9}{l_p^3} A_\sigma, &
A_\rho &\rightarrow \frac{RR_A}{l_p^3} A_\rho.
\end{aligned}
\end{equation}
The $2+1$ Yang-Mills theory action for the Matrix Membrane then
becomes 
\ben
S = \int d\tau \int_0^{2\pi\frac{l_p^3}{RR_9}} d\sigma
\int_0^{2\pi\frac{l_p^3}{RR_A}} d\rho~{\cal L},
\label{six}
\een
where
\bea
{\cal L} = \Tr & \{ & \frac{1}{2}[(D_\tau X^a)^2
- (D_\sigma X^a)^2 - e^{2Q\tau} (D_\rho X^a)^2]
+ \frac{1}{2(G_\text{YM}e^{Q\tau})^2}[F_{\sigma\tau}^2
+ e^{2Q\tau}(F_{\rho\tau}^2- F_{\rho\sigma}^2)] \nn \\
& - & 2\mu^2[(X^1)^2 + \cdots (X^6)^2 + 4 (X^7)^2] 
+\frac{(G_\text{YM}e^{Q\tau})^2}{4} [ X^a , X^b ]^2 \nn \\
& - & \frac{4\mu}{(G_\text{YM}e^{Q\tau})}e^{Q\tau} X^7~F_{\rho\sigma}
- 8\mu i (G_\text{YM}e^{Q\tau})X^7~[ X^5 , X^6 ]~ \},
\label{mmembraneaction}
\eea
with
\begin{equation}
G_\text{YM} = \sqrt{\frac{R}{R_A R_9}}.
\end{equation}
In the above $a,b = 1 \cdots 7$.
Using the relations (\ref{mtheoryconstants}) the radii of
the $\sigma$ and $\rho$ directions in (\ref{six}) become
${\tilde R}_9$ and ${\tilde R}_8$ of (\ref{one}) respectively.

The action~\eqref{mmembraneaction} has two important features:
\begin{enumerate}

\item Each factor of $G_\text{YM}$ is accompanied by a factor of
$e^{Q\tau}$.

\item Each factor of $\partial_\rho$ or $A_\rho$ is accompanied
by a factor of $e^{Q\tau}$.

\end{enumerate}

It is straightforward to see that one could rescale the fields
and the coordinates to write the action $S$ in the form
\ben
S = \frac{\mu}{G_\text{YM}^2} S(\mu=1, G_\text{YM}=1).
\een
In terms of these new coordinates, the range of $\sigma$ and
$\rho$ become
\ben \label{sizes}
0 \leq \sigma \leq \frac{\mu l_B^2}{R}, \qquad \qquad
0 \leq \rho \leq \frac{\mu l_B^2 g_B}{R}.
\een
The dimensionless coupling which controls the physics is then
\ben
\lambda = \frac{\mu}{G_\text{YM}^2} = \frac{\mu g_B l_B^4}{R R_B^2}.
\een
In the rest of the paper, however, we will stick to the choice of
coordinates and fields in which the action is given by  
(\ref{mmembraneaction}).

\section{Matrix Membrane Theory on Curved Space}

For the IIA big bang~\cite{Craps:2005wd,Das:2005vd}, the 
corresponding Matrix String Theory could be written as an ordinary gauge
theory---albeit with time-dependent masses~\cite{Das:2005vd}---on
Milne space, rather than Minkowski space.  We will see a similar phenomenon
occur here.

The first step is to perform a change of variables, $\rho = \rho' e^{Q \tau}$.
Although this inserts $\tau$-dependence into the range of $\rho'$, this is
natural in that the size of the $\rho$ direction
should be related to the string coupling---more precisely, the ratio of
the radii of $\rho$ and $\sigma$ circles is $g_s$,
\cf\ Eq.~\eqref{sizes}---which is
indeed $\tau$-dependent.
\
The resulting action is
\ben
S = \int d\tau \int_0^{2\pi\frac{l_p^3}{RR_9}} d\sigma
\int_0^{2\pi\frac{l_p^3}{RR_A} e^{-Q \tau}} d\rho'~{\mathcal L}',
\label{six'}
\een
where
\begin{multline}
{\mathcal L}' = \Tr  \left\{ - \frac{1}{2} e^{Q \tau} (D_\mu X^a)^2
- \frac{1}{4 G_\text{YM}^2} e^{-Q\tau} F_{\mu\nu}^2
 - 2\mu^2 e^{Q \tau} [(X^1)^2 + \cdots (X^6)^2 + 4 (X^7)^2] 
\right. \\ \left.
+\frac{G_\text{YM}^2}{4} e^{3 Q \tau} \com{X^a}{X^b}^2 
- \frac{4\mu}{G_\text{YM}} X^7~F_{\rho'\sigma}
- 8i \mu G_\text{YM}  e^{2 Q \tau} X^7~[ X^5 , X^6 ]~ \right\}.
\label{mmembraneaction'}
\end{multline}
Upon---as in~\cite{Das:2005vd}---defining $\mu(\tau) = \mu e^{-Q\tau}$,
this action---except for the funny, non-Lorentz-invariant
$X^7 F_{\rho\sigma}$ term, the consistency of which
will be explained in section~\ref{sec:review}
(see also~\cite{Michelson:2005iib})---can be
reinterpreted as a standard Yang-Mills Lagrangian on a space with metric
\begin{equation} \label{notMilne}
ds^2 = e^{2 Q \tau} \left[-d\tau^2 + d\sigma^2 + d\rho^{\prime 2}\right].
\end{equation}
That is,
\begin{multline}
S = \Tr \int d\tau \int_0^{2\pi\frac{l_p^3}{RR_9}} d\sigma
\int_0^{2\pi\frac{l_p^3}{RR_A} e^{-Q \tau}} d\rho'\, \sqrt{-g}\,
 \left\{ - \frac{1}{2} g^{\mu\nu} D_\mu X^a D_\nu X^a
\right. \\ \left.
- \frac{1}{4 G^2_\text{YM}} F_{\mu\nu} F^{\mu\nu}
 - 2\mu^2(\tau) [(X^1)^2 + \cdots (X^6)^2 + 4 (X^7)^2] 
\right. \\ \left.
+\frac{G_\text{YM}^2}{4}\com{X^a}{X^b}^2 
- \frac{4\mu(\tau)}{G_\text{YM}} e^{-2 Q \tau} X^7~F_{\rho'\sigma}
- 8 i \mu(\tau) G_\text{YM} X^7~\com{X^5}{X^6}~ \right\},
\end{multline}
with the metric~\eqref{notMilne}.

Interestingly, the metric~\eqref{notMilne} has a curvature singularity at
$\tau=-\infty$, which geodesics can reach in finite affine parameter.
In fact, this singularity is a worldvolume, spacelike big bang singularity.
This differs from the IIA big bang, in which this metric is that of the
flat two-dimensional Milne space.
That said, although the $\rho'$-direction appears to be shrinking
as $\tau\rightarrow -\infty$, its coordinate distance is growing as
$\tau\rightarrow \infty$. and thus the total 
physical size of the $\rho'$-direction is
actually constant in time.  (This is most easily seen by undoing the
coordinate transformation to $\rho'$, to obtain 
\hbox{$ds^2 = e^{2 Q \tau} \left(-d\tau^2 + d\sigma^2\right) 
+ \left(d\rho - Q \rho d\tau\right)^2$}.)
Nevertheless, this global picture does not change the local picture that gives
the big bang singularity at \hbox{$\tau=-\infty$}.

Since the worldvolume theory is nongravitational, the worldvolume big
bang cannot be resolved by quantum gravity (a.k.a.\ stringy) effects.
Thus, unlike the IIA big bang~\cite{Craps:2005wd}, one cannot attempt
to extrapolate through the $\tau=-\infty$ singularity to a pre-big bang
scenario.

\section{Behavior for $Q=0$} \label{sec:review}

Let us first review some aspects of the physics for usual
time independent \ppwave{s} following \cite{Michelson:2005iib}.
When the original IIB theory is weakly coupled, 
$g_B \ll 1$ with $\frac{\mu l_B^4}{R R_B^2} \sim O(1)$,
the effective coupling constant of this
YM theory becomes strong. The potential terms then restrict the fields
$X^a, F_{\mu\nu}$ to lie in the Cartan subalgebra and can be therefore
chosen to be diagonal. This also has the effect that the covariant
derivatives become ordinary derivatives.
We now perform a duality transformation followed by
a redefinition of fields:

\begin{enumerate}

\item Introduce an auxillary field $\phi$ and add a term to the
action $\frac{1}{2}\epsilon^{\mu\nu\lambda}\partial_\mu \phi
F_{\nu\lambda}$.

\item Integrate out the gauge fields to obtain a lagrangian 
density of the
form
\ben
{\cal L}' = -\frac{1}{2}[\sum_{a=1}^7
(\partial_\mu X^a)^2 + G_\text{YM}^2(\partial_\mu \phi)^2]
-2\mu^2[\sum_{i=1}^6 (X^i)^2 + 4 (X^7)^2] + 4
G_\text{YM}\mu~X^7\partial_\tau \phi.
\label{seven}
\een

\item Make a redefinition of the fields $(X^a,\phi) \rightarrow Y^I$
\begin{equation}
\begin{aligned}
X^i & = Y^i, \qquad i=1, \cdots, 6, \\
X^7 & = Y^7~\cos(2\mu\tau) + Y^8~\sin (2\mu\tau),  \\
G_\text{YM}\phi  & = -Y^7~\sin(2\mu\tau) + Y^8~\cos (2\mu\tau).
\end{aligned}
\label{eight}
\end{equation}
This leads to the lagrangian density
\ben
{\cal L}_{diag} =  -\frac{1}{2}\sum_{I=1}^8
(\partial_\mu Y^I)^2 -2\mu^2\sum_{I=1}^8 (Y^I)^2.
\label{nine}
\een
\end{enumerate}
In equations~\eqref{seven} and~\eqref{nine},
the {\em index\/} $\mu$ runs over all three
of the membrane worldvolume directions. However, in the 
$g_B \ll 1$ regime,
the size of the $\rho$ direction
shrinks to zero which implies that the $2+1$ dimensional theory
reduces to a $1+1$ dimensional theory. The lagrangian density
(\ref{nine}) is then precisely the (bosonic part of the)
light cone gauge fixed Green-Schwarz lagrangian density in 
the \ppwave\ background.
Allowed boundary
conditions lie in the conjugacy classes of the group which split the
action into pieces characterized by strips in $\sigma$ space whose
lengths add up to the total extent $J {\tilde R}_9$. This leads to
the worldsheet description of multiple light cone strings. The
fermionic terms also agree, as shown in \cite{Michelson:2005iib}. 
 
In the limit $\lambda \gg 1$ the Yang-Mills theory becomes
weakly coupled and classical solutions play a significant role.
These classical solutions include the {\em fuzzy ellipsoids} discussed
in \cite{Michelson:2004fh,Michelson:2005iib}.

Specifically, the potential of~\eqref{mtheoryaction} vanishes for the
static (taking $\mu$ to be constant) configuration
\begin{align}
X^5 &= 2 \sqrt{2} \frac{\mu l_p^3}{R} J^1, &
X^6 &= 2 \sqrt{2} \frac{\mu l_p^3}{R} J^2, &
X^7 &= 2 \frac{\mu l_p^3}{R} J^3,
\end{align}
where $J^a$ obey the SU(2) algebra, and the remaining matrices $X^i$ vanish.
(Actually, $X^8$ and $X^9$ need only commute with $X^{5,6,7}$,
corresponding to arbitrary positions of the fuzzy ellipsoids.)
By virtue of the SU(2) Casimir, on irreducible representations 
this configuration obeys,
\begin{equation}
(X^5)^2 + (X^6)^2 + 2 (X^7)^2
 = \text{constant}, 
\end{equation}
and so is ellipsoidal.
These vacua can be shown~\cite{Michelson:2004fh,Michelson:2005iib}
to preserve all 24 supercharges of the M-theory background.

\section{Shrinking ellipsoids}

In the presence of a $Q > 0$ the theory is again weakly coupled
near the big bang singularity at $\tau = -\infty$. Therefore,
in this regime there are fuzzy ellipsoids in addition to standard
strings. The time evolution of these fuzzy ellipsoids generalizes
that of the fuzzy spheres in IIA \ppwave{s} discussed in 
\cite{Das:2005vd}; for generic initial conditions, the size of these 
extended non-abelian configurations oscillate at early times and then
the ellipsoids degenerate as one evolves to $\tau = \infty$.
However, it turns out that
the ellipsoids do not shrink to zero size, but only to zero volume.
In particular, the solution depicted in Fig.~\ref{fig:degfuzzy}
exhibits exponential decay of what had been the major axes (along $X^5, X^6$)
of the ellipsoid, but the diagonal matrix
$X^7$ remains nonzero.  Thus, there
are (matrix) strings at late time, but they are ordinary and not giant
gravitons.

\FIGURE{
\includegraphics{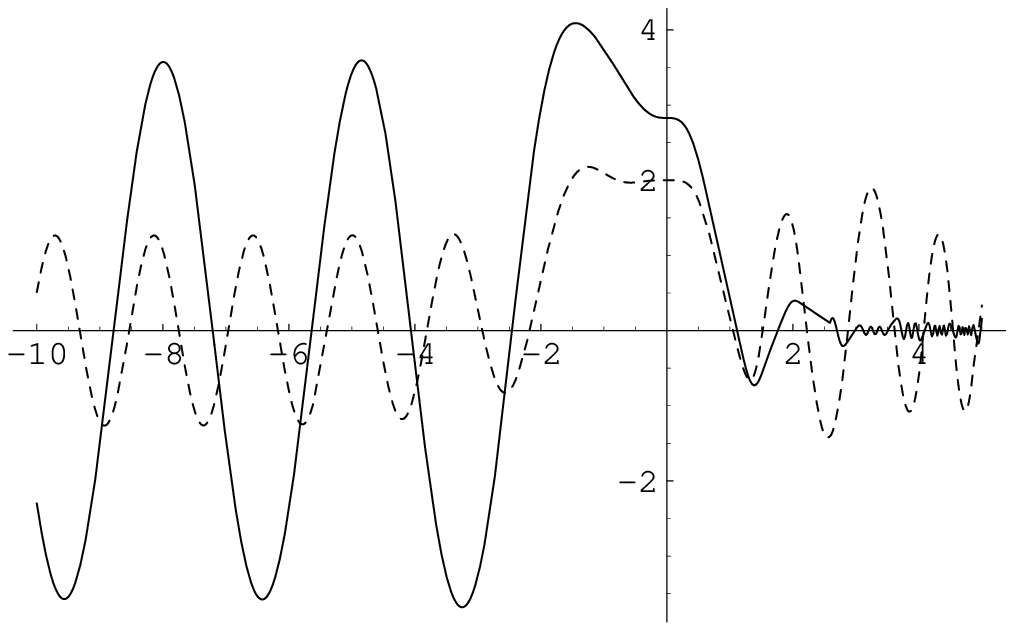}
\caption{The time evolution of the shape of the fuzzy ellipsoid,
for $A=1$ and a momentarily stationary fuzzy ellipsoid of the
(for $Q=0$) supersymmetric size.  The solid line is $R_1$, the size
in the $5,6$-directions, and the dashed line is $R_2$, the size in the
$7$-direction.  Thus, at late times, the fuzzy ellipsoid degenerates.
\label{fig:degfuzzy}}
}

Explicitly, consider the ansatz, with all other fields zero,
\begin{align}
X^5 &= \frac{\mu}{G_\text{YM}} R_1(\tau) J^1,
&
X^6 &= \frac{\mu}{G_\text{YM}} R_1(\tau) J^2,
&
X^7 &= \frac{\mu}{G_\text{YM}} R_2(\tau) J^3,
\end{align}
and set
\begin{align}
A &\equiv Q \sqrt{\frac{G_{\text{YM}}}{\mu^3}}, &
\tau &\equiv \sqrt{\frac{G_{\text{YM}}}{\mu^3}} t.
\end{align}
For $Q=0$, the equations of motion yield $R_1=2\sqrt{2}$ and $R_2=2$.
In general, the equations of motion imply the coupled differential equations
\begin{subequations}
\begin{align}
0 &= \ddot{R}_1 + 4 R_1 + e^{2 A t} R_1(R_1^2+R_2^2) - 8 e^{A t} R_1 R_2, \\
0 &= \ddot{R}_2 + 16 R_2 + 2 e^{2 A t} R_1^2 R_2 - 8 e^{A t} R_1^2,
\end{align}
\end{subequations}
where the dots denote $t$-derivatives.  Note that setting $R_1=0$ satisfies
the first equation, whereupon the second equation reduces to a harmonic
oscillator for $R_2$, which does not require that $R_2$ vanish.

Presumably a similar situation arises for IIA.  This was not seen
in~\cite{Das:2005vd} simply because the SO(3) symmetry made it natural to
start with a fuzzy sphere, whereupon the symmetry guaranteed that the
evolution preserved the spherical shape, but with exponential decay of the
radius.  The results here suggest that it
would be interesting to investigate nonspherical initial IIA configurations,
which might decay into (Matrix) strings.

\section{Particle Production and Initial States}

At late times, the $2+1$ dimensional theory is strongly coupled.
Therefore, as in the time-independent case discussed above, the fields
can be chosen to be diagonal. 
Exactly the same dualization and field redefinitions as
discussed above now lead to a $2+1$ dimensional lagrangian density
\ben
{\cal L}_{diag} = \frac{1}{2}[\sum_{I=1}^8
(\partial_\tau Y^I)^2 -(\partial_\sigma Y^I)^2
- e^{2Q\tau} (\partial_\rho Y^I)^2]
- 2\mu^2\sum_{I=1}^8 (Y^I)^2.
\label{ten}
\een
As expected, there is a factor of $e^{Q\tau}$ for each factor of
$\partial_\rho$. It is tempting to argue that as $\tau \rightarrow
\infty$ the Kaluza-Klein modes in the $\rho$ direction become
infinitely massive so that the theory becomes $1+1$ dimensional 
and exactly identical to the Green-Schwarz string action in this 
background. However, this is too hasty since we have a time-dependent
background here and energetic arguments do not apply.

Since the size of the $\rho$ direction is given by ${\tilde{R_8}}$
given in (\ref{one}), 
the mass scale associated with these Kaluza-Klein modes is
$M_{KK} \sim {R \over g_B l_B^2}$ while the mass scale associated with
the coupling is $G_{YM}^2$ which is given in (\ref{two}).
Therefore when $R_B \gg l_B$ the KK modes are much lighter than the 
Yang-Mills scale. In our present time-dependent context, these scales
become time-dependent and it follows from the coupling and the
$\partial_\rho$ terms in (\ref{mmembraneaction}) that the KK
modes are expected to decouple much {\em later} than the time when the
non-abelian excitations decouple. Therefore, there is a regime where
we can ignore the non-abelian excitations, but cannot ignore the KK
modes. In this regime, the Matrix Membrane lagrangian density is 
given by (\ref{ten}). In the following we will assume that the time
interval for which (\ref{ten}) is valid is long enough
to be well approximated by the entire interval ($-\infty$,$\infty$).


To determine the fate of these KK modes we need to find the
modes of the field $Y^I$.
The mode functions
which are positive frequency at
early times are
\ben
\varphi^{(\text{in})}_{m,n} = 
\{ {\frac{R}{8 \pi^2 l_B^4 g_B}} \}^{1/2}~
\Gamma (1-i\omega_m / Q)~e^{i(\frac{mR}{l_B^2}\sigma
+\frac{n R}{g_B l_B^2}\rho)}~J_{-i\frac{\omega_m}{Q}}
(\kappa_n e^{Q\tau}),
\label{inmodes}
\een
where
\ben
\omega_m^2 = 4\mu^2 + \frac{m^2R^2}{l_B^4}, \qquad
\kappa_n = \frac{nR}{Qg_B l_B^2},
\een
while those which are appropriate at late times are
\ben
\varphi^{(\text{out})}_{m,n} = 
\{ \frac{R}{16 \pi l_B^4 g_B Q}\}^{1/2}~
e^{i(\frac{mR}{l_B^2}\sigma
+\frac{n R}{g_B l_B^2}\rho)}~H^{(2)}_{-i\frac{\omega_m}{Q}}
(\kappa_n e^{Q\tau}).
\label{outmodes}
\een
The problem is of course equivalent to that of a bunch of
two dimensional scalar fields 
with time-dependent masses and it is well known that such
time-dependent masses lead to particle
production or depletion \cite{Strominger:2002pc,Birrell:1982ix,
Tanaka:1996cz}. 
Because of standard relations between the Hankel function $H^{(2)}_\nu(z)$
and the Bessel function $J_{\nu}(z)$ there is a non-trivial Bogoliubov
transformation between these modes which imply that the vacua defined
by the in and out modes are not equivalent. Indeed one has
\ben
\varphi^{(\text{out})}_{m,n} = \alpha_m \varphi^{(\text{in})}_{m,n} 
+ \beta_m 
(\varphi^{(\text{in})}_{-m,-n})^\star.
\een
where
\ben
\alpha_m = \{\frac{Q}{2\pi\omega_m} \}^{1/2}~
\Gamma(1+i\omega_m / Q)~e^{\frac{\pi\omega_m}{2Q}}, \qquad
\beta_m = - \{ \frac{Q}{2\pi\omega_m} \}^{1/2}
\Gamma(1-i\omega_m / Q)~e^{-\frac{\pi\omega_m}{2Q}}.
\een
This means that the out vacuum $\ket{0}_\text{out}$ is a squeezed state of the
``in'' particles
\ben
\ket{0}_\text{out} = \prod_{n,m} \
\{ (1-\abs{\gamma_m}^2)^{1/4}~\exp[\frac{1}{2}\gamma_m^\star 
a^{\dagger I,(\text{in})}_{m,n}~a^{\dagger I,(\text{in})}_{-m,-n}] \} 
~\ket{0}_\text{in},
\label{inout}
\een
where $a^{I,(\text{in})}_{m,n}$ is the annihilation operator of the KK
mode labeled by $n$ with $m$ units of momentum in the $\sigma$
direction. Here
\ben
\gamma_m = \frac{\beta^\star_m}{\alpha_m} = -ie^{-\frac{\pi\omega_m}{Q}}.
\een
There is of course a similar relationship which expresses the in vacuum as a
squeezed state of the out particles.

In the present context, the relation (\ref{inout}) means that {\em if
we require that the final state at late times does not contain any of
the KK modes, the initial state must be a squeezed state of these
modes}. The occupation number of the in modes in this state is thermal
\ben
_\text{out}\bra{0}a^{\dagger I,(\text{in})}_{m,n} a^{I,(\text{in})}_{m
,n} \ket{0}_\text{out} = \frac{1}{e^{\frac{2\pi\omega_m}{Q}}-1}.
\een
Note that the Bogoliubov
coefficients and number densitites depend only on $m$ for all
$n \neq 0$. This follows from the fact that $n$-dependence may
be removed by shifting the time $\tau$ by $\log (\kappa_n)$.
However, the modes with $n=0$ need special treatment.  Indeed, in the $n
\rightarrow 0$ limit the ``in'' modes (\ref{inmodes}) go over to
standard positive frequency modes of the form $e^{-i\omega_m \tau}$ as
expected. In this limit, however, the out modes (\ref{outmodes})
contain both positive and negative frequencies. This is of course a
wrong choice, since for these $n=0$ modes there is no difference
between ``in'' and ``out'' states.  In fact, the ``out'' modes
(\ref{outmodes}) have been chosen by considering an appropriate large
time property for {\em nonzero n} and do not apply for $n=0$.
In other words, the squeezed state (\ref{inout}) contains only
the $n \neq 0$ modes.

The phenomenon we described above of course occurs even when $\mu =0$, in 
which case the IIB string frame metric is flat. In this case,
it is useful to consider the lagrangian as a sum of the lagrangian
densities of an infinite number of 1+1 dimensional scalar and fermion
fields with time-dependent masses 
\ben
M_n^2 (\tau) = [ \frac{nR}{g_B l_B^2} ]^2~e^{2Q\tau}.
\een
The resulting action may be viewed as that of fields
with time {\em independent} masses (and their fermionic partners) in a 
Milne universe.
In a way similar to the IIA background of \cite{Craps:2005wd}
the theory for each $n$ has a conserved supercurrent. However 
the periodicity in $\sigma$ leads to a breaking of supersymmetry
due to boundary conditions. Indeed if $\sigma$ were non-compact,
one could make a coordinate transformation to ``Minkowski''
coordinates in which the supersymmetry is obvious. The ``Minkowski''
vacuum is supersymmetric and there is no particle production.
The periodicity of $\sigma$ destroys this property, and there is
no supersymmetric vacuum since there is no supersymmetry. However,
as $\tau \rightarrow \infty$ the periodicity of $\sigma$ becomes
less important and constant $\tau$ slices approach slices of 
constant ``Minkowski'' time. Indeed the ``out'' vacuum defined above
is identical to the ``Minkowski'' vacuum: the positive frequency
modes defined with respect to these out modes are also positive
frequency with respect to the ``Minkowski'' modes \cite{Tanaka:1996cz}.

The operators $a^I_{m,n}$ in fact create states of $(p.q)$ strings 
in the original Type IIB theory
\cite{Banks:1996my}. To see this, let us recall how the light cone IIB 
fundamental string states arise from the $n=0$ modes of the Matrix
Membrane. In this sector the action is exactly the Green-Schwarz
action. The oscillators $a^{\dagger I}_{m,0}$ defined above are
in fact the world sheet oscillators and create excited states of a
string. The gauge invariance of the theory allows nontrivial boundary
conditions, so that $m$ defined above can be fractional. Equivalently
the boundary conditions are characterized by conjugacy classes of the
gauge group. The longest cycle corresponds to a single string whose
$\sigma$ coordinate has an extent of $2\pi J {l_B^2 \over R}$ which is
the same as $2\pi l_B^2 p_-$ as it should be in the light cone gauge.
Shorter cycles lead to multiple strings - the sum of the lengths of
the strings is always $2\pi l_B^2 p_-$, so that there could be at most
$J$ strings. 
Note that $m$ is the momentum in the $\sigma$ direction:
a state with {\em net} momentum in
the $\sigma$ direction in fact corresponds to a fundamental IIB string
wound in the $x^-$ direction. This may be easily seen from the chain
of dualities which led to the Matrix Membrane. 

As shown in \cite{Banks:1996my}, following the arguments of
\cite{Schwarz:1995dk,sM,Aspinwall:1995fw}, $SL(2,\ZZ)$ transformations on
the torus on which the Yang-Mills theory lives become the $SL(2,\ZZ)$
transformations which relate $(p,q)$ strings in the original IIB theory.
In particular the oscillators $a^I_{0,n}$ create states of a 
D-string.

The squeezed state (\ref{inout}) is therefore a superposition of 
excited states of these
$(p,q)$ strings. The number of such strings depends on the choice of
the conjugacy classes characterizing boundary conditions.  Since each
$(m,n)$ quantum number is accompanied by a partner with $(-m,-n)$ this
state does not carry any F-string or D-string winding number. Finally
this squeezed state contains only $n \neq 0$ modes, \ie\ they do not
contain the states of a pure F-string.
We therefore conclude that in this toy model the initial state has to
be chosen as a special squeezed state of {\em unwound} $(p,q)$ strings
near the big bang to ensure that the late time spectrum contains only
perturbative strings.
It is interesting that this toy model of
cosmology can address the issue of initial conditions.

In the above discussion we have ignored the effect of D-string interactions.
In fact when $g_B \ll 1$ the pure D-strings described above are strongly
coupled and all excited states rapidly decay into supergravity modes.
However states with $n \neq 0$ which 
are ``almost'' F-strings could be relatively long lived. For such states
the above results based on free strings will continue to be relevant.%
\footnote{Previous work on string pair production includes~\cite{sp1,%
sp2,sp3,sp4,sp5,sp6,membranepair}.}

Finally, in this paper we have not considered the possible generation
of potentials for the fields due to quantum effects. In the absence of
\ppwave{s} this indeed happens \cite{Li:2005ai,Craps:2006xq} and one
would expect that the same would be true in the presence of \ppwave{s}.
However, as found in \cite{Li:2005ai,Craps:2006xq} and emphasized in
\cite{Craps:2006xq} the potential vanishes at early times (as expected)
{\em as well as at late times}. This suggests that standard
perturbative string physics is indeed recovered at late times. We have
not yet performed a similar analysis in the \ppwave\ background, but we
expect similar results to hold. In our discussion of particle
production, however, we implicitly assumed that the potential vanishes
at an intermediate time where the non-abelian excitations have
decoupled, but the Kaluza-Klein modes have not. This requires a
detailed investigation. The presence of a potential will certainly
change the details of particle production. However we expect that the
basic fact that $(p,q)$ strings are produced to still hold. 

It would be interesting to explore the meaning of the supergravity
background in the holographic dual in terms of the 3+1 dimensional
gauge theory. This seems to require a deformation of $AdS_5 \times S^5$
which, in the Penrose limit, would become the solution of this paper.
In \cite{chuho,dmnt}
a large class of time-dependent deformations of $AdS_5 \times S^5$
have been found for which there is a natural proposal for the dual
gauge theory. Although this class does not include the one we are looking for,
further investigations along these lines might lead to the answer.

\acknowledgments
We would like to thank K. Narayan and 
Sandip Trivedi for numerous discussions and collaboration at early
stages. We also thank Samir Mathur, Alfred Shapere and Xinkai Wu for
helpful conversations. S.R.D. would like to thank Tata Institute of
Fundamental Research, Mumbai for hospitality. 
This work was supported in part by a National
Science Foundation grant No.\ PHY-0244811 and a Department of Energy
contract
\#DE-FG01-00ER45832.

\end{document}